\documentclass{aastex}
\usepackage{graphicx,epsfig}
\usepackage{emulateapj5}
\makeatletter

\newenvironment{inlinefigure}{%
\def\@captype{figure}%
\noindent\begin{minipage}{0.999\linewidth}\begin{center}}
{\end{center}\end{minipage}\smallskip}
\makeatother

\begin{document}
\title{Discovery of the largest known lensed images formed by a
critically convergent lensing cluster}
\def\HI{\hbox{H~$\scriptstyle\rm I\ $}}
\def\arcs{$''$}
\def\arcm{$'$}
\def\sfrd{\,{\rm M_\odot\,yr^{-1}\,Mpc^{-3}}}
\def\emunits{\,{\rm ergs\,s^{-1}\,Hz^{-1}\,Mpc^{-3}}}
\def\be{\begin{equation}}
\def\ee{\end{equation}}

\author{Adi Zitrin \& Tom Broadhurst}
\affil{School of Physics and Astronomy, Tel Aviv University, Israel}

\shorttitle{A Critically Convergent Lens Forming The Largest Lensed Images}
\shortauthors{Zitrin \& Broadhurst}

\slugcomment{Accepted to the Astrophysical Journal Letters}

\begin{abstract}

We identify the largest known lensed images of a single spiral galaxy,
lying close to the centre of the distant cluster MACS J1149.5+2223
($z=0.544$). These images cover a total area of $\simeq
150~\Box\arcsec$ and are magnified $\simeq 200$ times. Unusually,
there is very little image distortion implying the central mass
distribution is almost uniform over a wide area ($r\simeq200~kpc$)
with a surface density equal to the critical density for lensing,
corresponding to maximal lens magnification. Many fainter
multiply-lensed galaxies are also uncovered by our model, outlining a
very large tangential critical curve, of radius $r\simeq 170~kpc$,
posing a potential challenge for the standard
LCDM-Cosmology. Because of the uniform central mass distribution a
particularly clean measurement of the mass of the brightest cluster
galaxy is possible here, for which we infer stars contribute most of
the mass within a limiting radius of $\simeq 30~kpc$, with a
mass-to-light ratio of $M/L_{B}\simeq 4.5(M/L)_{\odot}$. This cluster
with its uniform and central mass distribution acts analogously to a
regular magnifying glass, converging light without distorting the
images, resulting in the most powerful lens yet discovered for
accessing the faint high-$z$ Universe.

\end{abstract}

\section{Introduction}

Multiply-lensed images of distant galaxies are commonly seen near the
centres of distant galaxy clusters. Thin tangentially stretched
images, including giant arcs, follow an approximate ``Einstein ring",
interior to which radially directed arcs are sometimes found, pointing
towards the centre of mass. The formation of such strongly lensed
images requires the projected mass density to exceed a critical value,
$\Sigma_{crit}$, inside the Einstein radius, given by fundamental
constants and the angular-diameter distance of the lens, $d_l$, and of
the source $d_s$, and their separation, $d_{ls}$, such that
$\Sigma_{crit}={c^2\over{4\pi G}}{D_{s}\over{D_l D_{ls}}}$. This mass
density is a little less than $1~g/cm^2$ for lenses at intermediate
distances and comprises mainly dark matter (DM) whose nature is still
unknown, with only a $\sim$ 5-15\% contribution from gaseous baryonic
material (Biviano \& Salucci, 2006, Umetsu et al., 2009).

Fine examples of lensing by galaxy clusters are found at intermediate
redshifts, including the most distant cluster discovered by Zwicky,
Cl0024+17 ($z=0.39$), and one of the richest clusters discovered by
Abell, A1689 ($z=0.18$), with the largest known Einstein ring, $r
\simeq 45"$ (Broadhurst et al., 2005). For such clusters many tens of multiply lensed
images have been identified in deep Hubble images (Broadhurst et al.,
2005, Halkola et al., 2008, Limousin et al., 2008, Zitrin et al.,
2009a,b), leading to accurately measured central surface mass
distributions.  The Einstein radii of these massive clusters are found
to be larger than predicted in the context of the standard LCDM
cosmological model (Broadhurst \& Barkana, 2008, Sadeh \& Rephaeli,
2008), based on the ``Millenium'' simulation (Springel et al.,
2005). This discrepancy is empirically supported by the surprisingly
concentrated mass profiles measured for such clusters, when combining
the inner strong lensing with the outer weak lensing signal (Gavazzi
et al., 2003, Broadhurst et al., 2008, Limousin et al., 2008, Umetsu
et al., 2009, Zitrin et al., 2009a, Oguri et al., 2009, Donnarumma et
al., 2009) boosting the critical radius at fixed virial mass.
The abundance of giant arcs may help constrain the total lensing
cross section and is variously claimed to be at odds with standard
LCDM (Bartelmann et al., 1998), though recent work favours the
consensus (Dalal et al., 2004, Wambsganss et al., 2004, Horesh et al.,
2005, Hennawi et al., 2007) here the selection effects are considerable and
more thorough surveys should help (Hennawi et al. 2007).

The magnification generated by massive clusters has consistently led
to the discovery of the highest redshift galaxies (Ebbels et al.,
1995, Franx et al., 1997, Frye \& Broadhurst, 1998, Bouwens et al.,
2004, Kneib et al., 2004, Zheng et al., 2009), with the current record
standing at $z\simeq7.6$ for a galaxy behind A1689 (Bradley et al.,
2008) and magnified by nearly a factor of $\sim 10$. Although lens
magnification, $\mu$, reduces the accessible area of the source plane
by $1/\mu$, it enhances the flux of faint galaxies by $\mu$, with a
net positive effect for the most distant galaxies lying on the steep
exponential tail of the luminosity function (Broadhurst, Taylor \&
Peacock, 1995, Bradley et al., 2008). Lensing provides additional
spatial resolution by stretching images, producing spatially resolved
details and evidence that outflowing gas is widespread at $z>4$ (Frye,
Broadhurst \& Ben\'itez, 2002).

With the goal of discovering high redshift galaxies and to better
define the mass profiles of galaxy clusters in general, we have
combined data for a sample of well studied clusters. In this process
we have uncovered the unusual lensing properties of MACS J1149.5+2223
($z$=0.544), a cluster originally identified in the highly complete
sample of the most X-ray luminous clusters in the Universe (Ebeling et
al., 2007). In \S2 we describe the observations, in \S3 we describe
the lensing analysis and results, in \S4 we address the BCG, and in
\S5 discuss our conclusions. Throughout the paper we adopt the
standard cosmology ($\Omega_{\rm m0}=0.3$, $\Omega_{\Lambda0}=0.7$,
$h=0.7$). Accordingly, one arcsecond corresponds to 6.4 kpc$/h_{70}$
at the redshift of this cluster. The reference center of our analysis
is fixed at the center of the cD galaxy: RA = 11:49:35.70, Dec =
+22:23:54.8 (J2000.0).

\section{Observations}
The central region of the luminous X-ray cluster MACS J1149.5+2223
($z=0.544$) has been imaged in April 2004 and in May 2006, with the
Wide Field Channel (WFC) of the ACS installed on HST. Integration
times of $\sim 4500s$ were obtained through the F555W and the F814W
filters. We retrieved these images (PI: Ebeling, proposal ID: 9722)
found in the Hubble Legacy Archive. Several large blue spiral galaxy
images are clearly visible near the central brightest cluster galaxy
(Figures 1-3). On closer inspection, individual HII regions and spiral arms
are repeated very clearly in images 1.1 and 1.2, demonstrating beyond
question that these are images of the same source, even though they do
not appear as thin distorted arcs. The other central spiral galaxy
images seen in Figure~1-3 are also images of the same source but with
differing mirror symmetry (parity), as we show below.  Image 1.2 is
the largest, covering an area of $\simeq 55~\Box\arcsec $, and in
total $\gtrsim 150~\Box\arcsec$ is subtended by all the images of this
source, several times greater than the largest giant arc known
(Soucail et al., 1987).

 Many other faint lensed galaxies are also visible, generally at
 larger distances from the centre (marked in Figure 2) indicating that
 they lie at higher redshift and most of which we have been able to
 securely identify as sets of multiply-lensed background galaxies
 as detailed below.

\begin{inlinefigure}
\vspace{1cm}
 \begin{center}
  \epsfig{figure=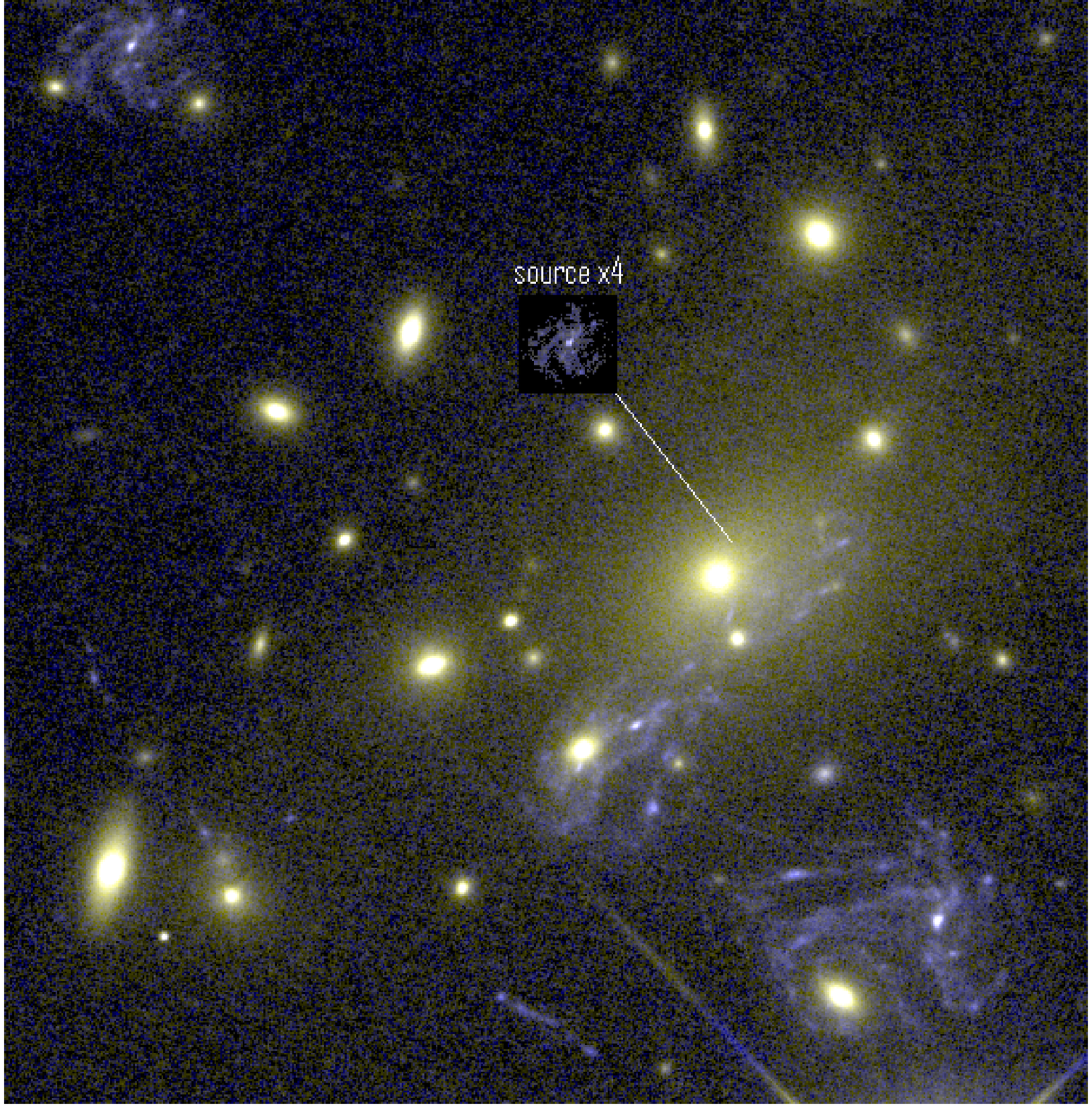,width=3.2in}
 \end{center}
\caption{The central region of the cluster MACS J1149.5+2223 ($z=0.544$) imaged
with Hubble/ACS in V and I-bands, showing the several large, well
resolved, blue spiral galaxy images centered on the brightest red
cluster member galaxy (cD). The small delensed source, zoomed-in $\times$4
to show internal details, is inset near its predicted location
demonstrating the very large magnifications generated by the
central gravitational field of this cluster.}
\end{inlinefigure}

\section{Lensing Analysis}

 We apply our well tested approach to lens modelling, which we have
 applied successfully to A1689 and Cl0024, uncovering unprecedentedly
 large numbers of multiply-lensed images (Broadhurst et al., 2005, Zitrin
 et al., 2009a). The full details of this approach can be found in these
 papers. Briefly, the basic assumption adopted is that mass
 approximately traces light, so that the photometry of the red cluster
 member galaxies is the starting point for our model.

 Cluster member galaxies are identified as lying close to the cluster
 sequence by the photometry provided in the Hubble Legacy Archive.  We
 approximate the large scale distribution of matter by assigning a
 power-law mass profile to each galaxy, the sum of which is then
 smoothed. The degree of smoothing and the index of the power-law are
 the most important free parameters. A worthwhile improvement in
 fitting the location of the lensed images is generally found by
 expanding to first order the gravitational potential of the smooth
 component, equivalent to a coherent shear, where the direction of the
 shear and its amplitude are free, allowing for some flexibility in
 the relation between the distribution of DM and the distribution of
 galaxies which cannot be expected to trace each other in detail. The
 total deflection field $\vec\alpha_T(\vec\theta)$, consists of the
 galaxy component, $\vec{\alpha}_{gal}(\vec\theta)$, scaled by a
 factor $K_{gal}$, the cluster DM component
 $\vec\alpha_{DM}(\vec\theta)$, scaled by (1-$K_{gal}$), and the
 external shear component $\vec\alpha_{ex}(\vec\theta)$:

\begin{equation}
\label{defTotAdd}
\vec\alpha_T(\vec\theta)= K_{gal} \vec{\alpha}_{gal}(\vec\theta)
+(1-K_{gal}) \vec\alpha_{DM}(\vec\theta)
+\vec\alpha_{ex}(\vec\theta),
\end{equation}
where the deflection field at position $\vec\theta_m$
due to the external shear,
$\vec{\alpha}_{ex}(\vec\theta_m)=(\alpha_{ex,x},\alpha_{ex,y})$,
is given by:
\begin{equation}
\label{shearsx}
\alpha_{ex,x}(\vec\theta_m)
= |\gamma| \cos(2\phi_{\gamma})\Delta x_m
+ |\gamma| \sin(2\phi_{\gamma})\Delta y_m,
\end{equation}
\begin{equation}
\label{shearsy}
\alpha_{ex,y}(\vec\theta_m)
= |\gamma| \sin(2\phi_{\gamma})\Delta x_m -
  |\gamma| \cos(2\phi_{\gamma})\Delta y_m,
\end{equation}

where $(\Delta x_m,\Delta y_m)$ is the displacement vector of the
position $\vec\theta_m$ with respect to a fiducial reference position,
which we take as the lower-left pixel position $(1,1)$, and
$\phi_{\gamma}$ is the position angle of the spin-2 external
gravitational shear measured anti-clockwise from the $x$-axis.
The normalisation of the model and the relative scaling of the
smooth DM component versus the galaxy contribution brings the
total number of free parameters in the model to 6.

We lens all well detected candidate lensed galaxies back to the
source plane using the derived deflection field, and then relens this
source plane to predict the detailed appearance and location of
additional counter images, which may then be identified in the data by
morphology, internal structure and colour. The fit is assessed by the
RMS uncertainty in the image plane:

\begin{equation} \label{RMS}
RMS_{images}^{2}=\sum_{i} ((x_{i}^{'}-x_{i})^2 + (y_{i}^{'}-y_{i})^2) ~/ ~N_{ima
ges},
\end{equation}
where $x_{i}^{'}$ and $y_{i}^{'}$ are the locations given by the
model, and $x_{i}$ and $y_{i}$ are the real images location, and the
sum is over all $N_{images}$ images. The best-fit solution is
unique in this context, and the uncertainties are determined by the
location of predicted images in the image plane.

Importantly, this image-plane minimisation does not suffer from the
well known bias involved with source plane minimization, where
solutions are biassed by minimal scatter towards shallow mass profiles
with correspondingly higher magnification. The model is successively
refined as additional sets of multiple images are identified and then
incorporated to improve the model (Zitrin et al., 2009a).

\begin{inlinefigure}
\vspace{1cm}
 \begin{center}
 \epsfig{figure=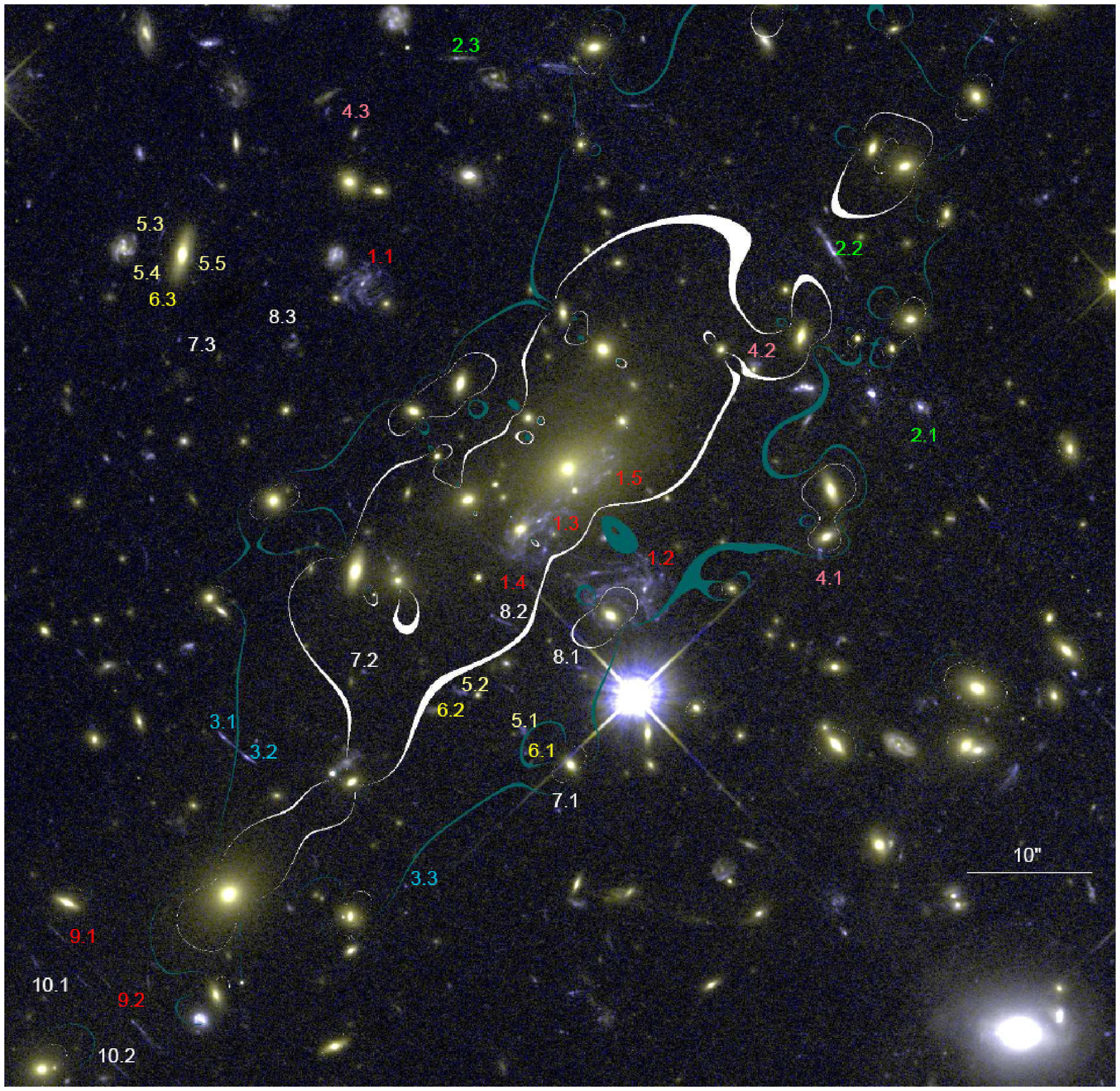,width=3.2in}
 \end{center}
\caption{Large scale view of the multiply lensed galaxies identified by our
model. In addition to the large spiral galaxy system~1, many other
fainter sets of multiply lensed galaxies are uncovered by our
model. The white curve overlaid shows the tangential critical curve
corresponding to the distance of system~1. The larger critical curve
overlaid in blue corresponds to the average distance of the fainter
systems, passing through close pairs of lensed images in systems 2 and
3. This large scale elongated ``Einstein ring'' encloses a very large
critically lensed region equivalent to $170~kpc$ in radius. For this cluster one arcsecond
corresponds to 6.4 kpc$/h_{70}$, with the standard cosmology.}
\end{inlinefigure}

\begin{inlinefigure}
\vspace{1cm}
 \begin{center}
  \epsfig{figure=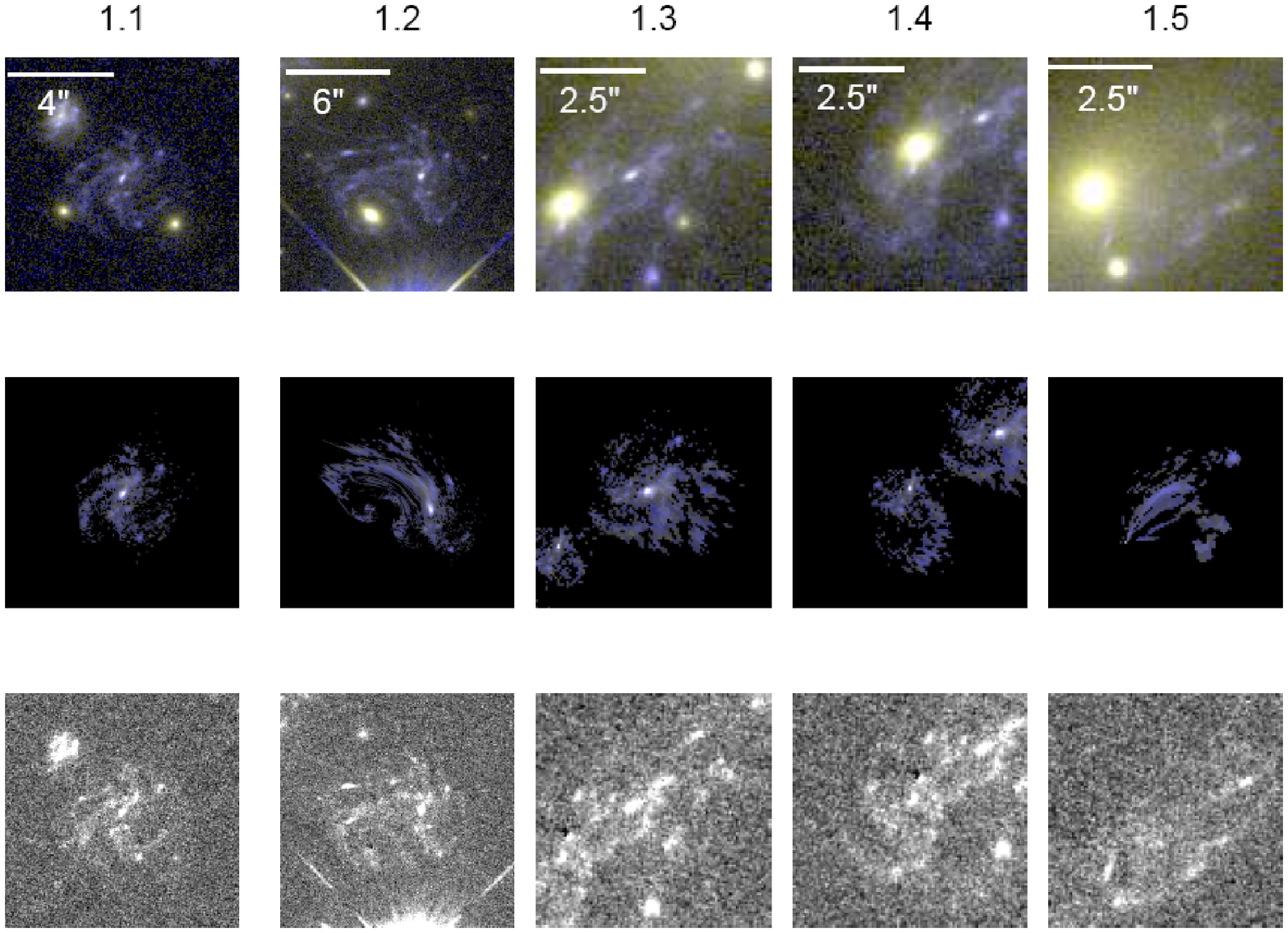,width=3.2in}
 \end{center}
\caption{The observed images are shown with (top row) and without (bottom row)
the light of the cluster member galaxies, and compared with our model
generated images (middle row). For the model we take as input the
pixels belonging to 1.1, the least distorted and cleanest image in
system 1, and delens these pixels back to the source plane (see inset in
figure~1) and then relens the source plane to generate counter
images. It is clear that our modelling is successful in demonstrating
the multiply-lensed origin of all 5 observed images, corresponding to a
single distant spiral galaxy.}
\end{inlinefigure}

\begin{inlinefigure}
\vspace{1cm}
 \begin{center}
  \epsfig{figure=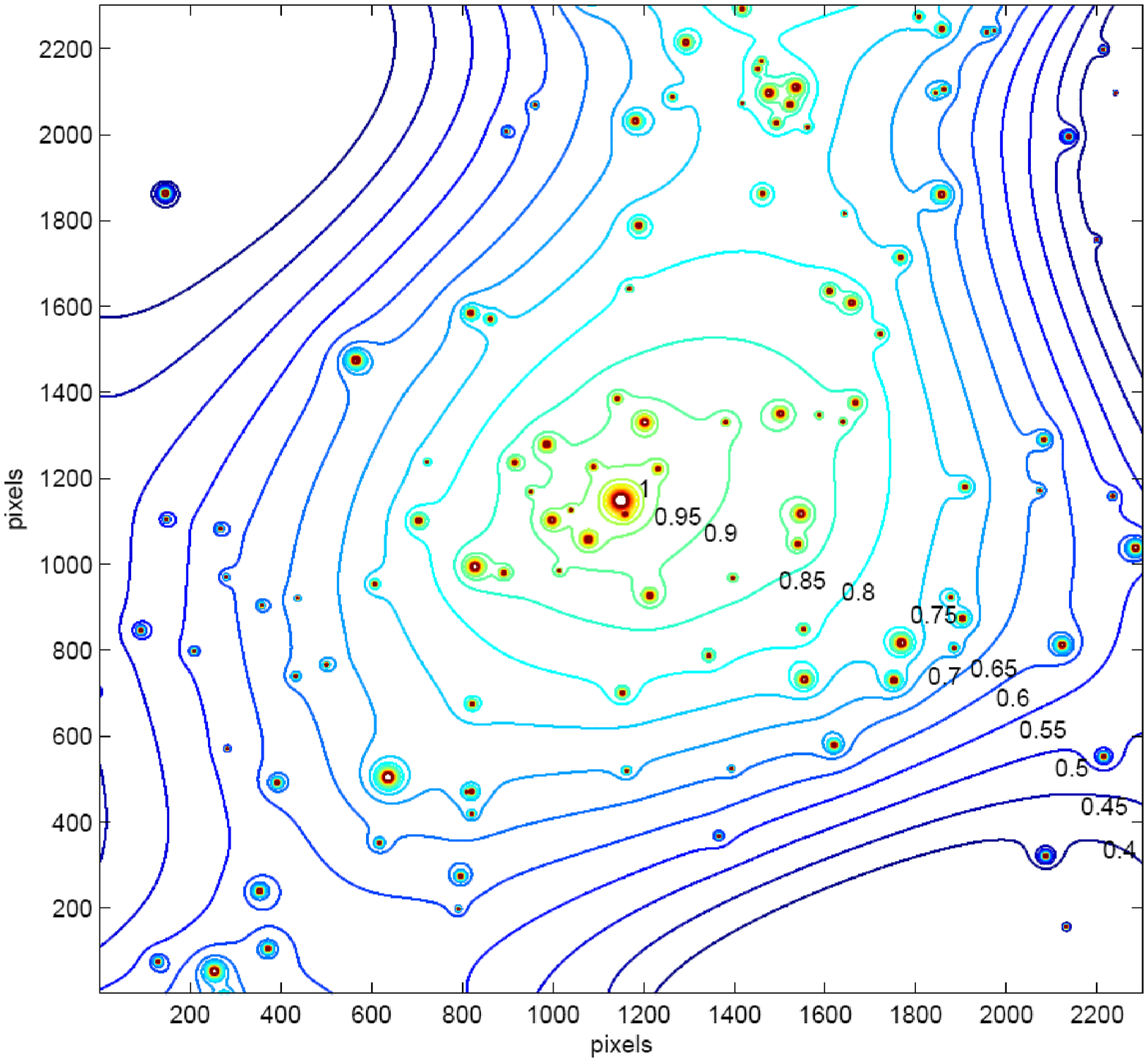,width=3.2in}
 \end{center}
\caption{2D surface mass distribution ($\kappa$), in units of the critical density. Contours are shown in linear units, derived from the mass model constrained using 33 multiply lensed images seen in Figure 2. Note, the central mass distribution is shallow, and rounder in shape than the critical curves.}
\end{inlinefigure}

The derived surface mass distribution is found to be very nearly
uniform within the central $\simeq 200~kpc$ (Figures 4-5), with very
little uncertainty, as must be expected given the very large and
undistorted images observed. The value of the uniform surface mass in
this central region is the critical value for generating multiple
images, about $\simeq$ 0.50~$g/cm^2$ at our estimated
redshift for the source, see below.

The total magnification of the spiral galaxy we derive is about, $\simeq
200$, when summed over all five images, forming the largest known
images of any lensed source, and is independent of the unknown source
redshift, and given by the ratio of the area of images divided by the
area subtended by the deprojected source. This can be appreciated in
Figure~1 where we plot for comparison the unlensed source and its
modelled location on the sky, whose diameter is estimated to be
$\simeq 0.9"$, typical of unlensed spiral galaxies at intermediate
redshifts.

The critical curve corresponding to the spiral system is shown in
Figure 2, together with the larger Einstein radius derived for the
fainter galaxies, which lie at higher redshift, typically in the range
$z\sim 1.5-2.5$, based on other massive clusters where redshift
measurements have been made or photometric redshifts obtained with
more colour information (Broadhurst et al., 2005, Zitrin et al.,
2009a). From this we can roughly calculate the relative lensing
distance of the spiral system~1, from the ratio $d_{ls}/d_s$. For
a mean background redshift of $z=1.5$, the
spiral system is at $z\simeq1.2$, and at the other extreme for a
background redshift of $z=2.5$ the maximum predicted redshift of the spiral is
$z\simeq2$. Assuming an average background depth of $z\simeq2\pm0.5$ yields
$z_{spiral}=1.5^{+0.5}_{-0.3}$, due to the shape of the
$d_{ls}/d_s$ growth as a function of source redshift for a cluster at $z=0.544$.
The shape, location and magnification of the critical curves are unaffected by this
uncertainty. The tangential critical curve is elongated, with a
major axis of $\sim 80''$ at a redshift of $\sim 2$ (Figure 2),
reflecting the somewhat elongated distribution of matter (Figure
4). Note, the mass distribution is more symmetric than the shape of
the tangential critical curve, the form of which is sensitive to
asymmetry.  The circular equivalent Einstein radius contained within
the critical area is $\simeq 27"$ for faint sources at $z\sim 2$,
corresponding to the outer tangential critical curve drawn in Figure 2
(light blue curve), which in physical units is $170\pm 20~kpc$ at the
distance of the cluster, where the uncertainty is dominated by the
uncertain mean redshift of the background galaxies. Note that because
of the weak redshift dependence of lensing distance, $d_{ls}/d_s$, the
uncertain redshifts are only a minor source of uncertainty when
deriving physical scales.

\begin{inlinefigure}
\vspace{1cm}
 \begin{center}
 \epsfig{figure=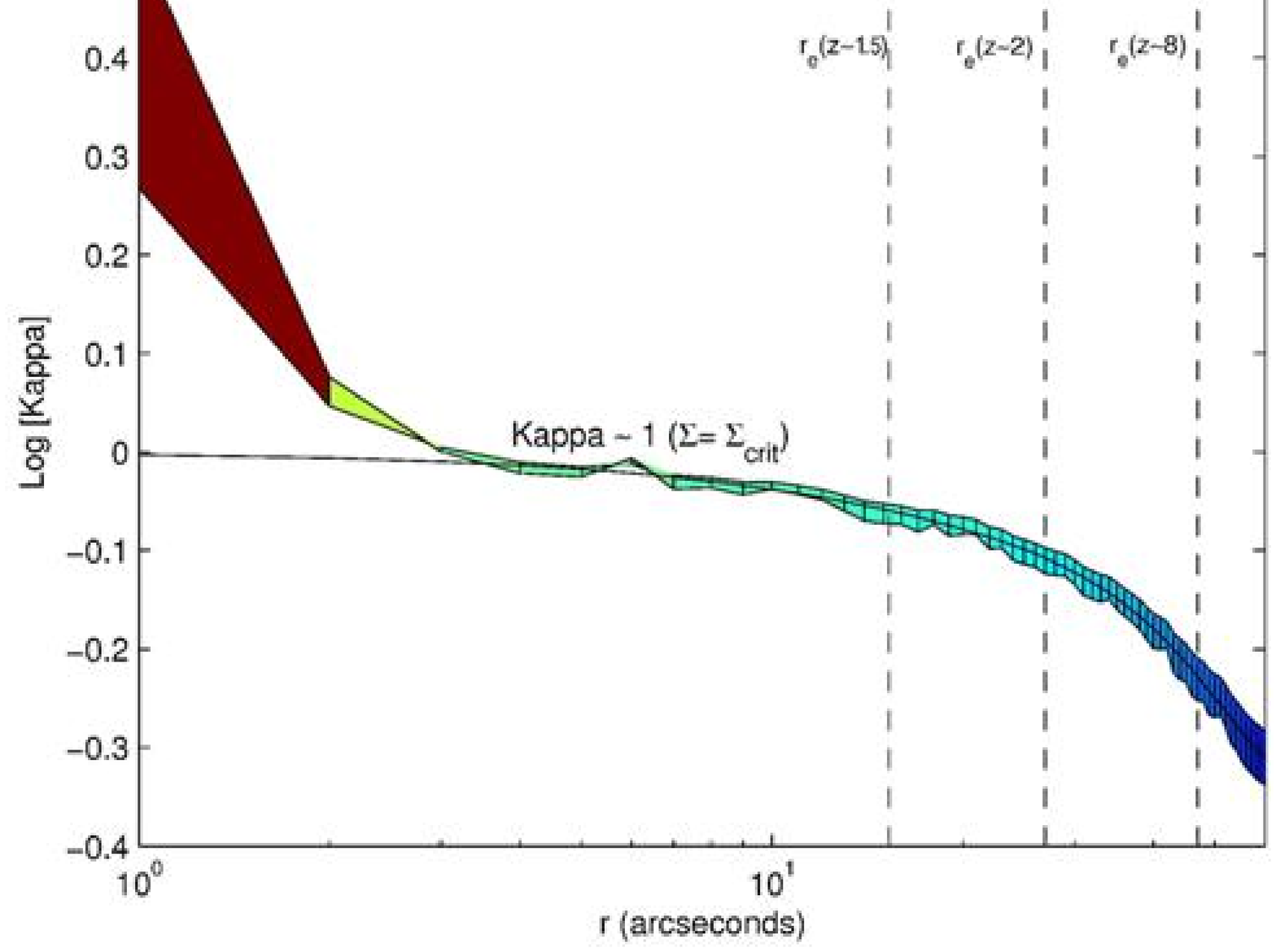,width=3.2in}
 \end{center}
\caption{Radial surface mass density profile, $\kappa(r)$, in units of the
critical surface density, i.e.  $\kappa(r)=\Sigma(r)/\Sigma_{crit}$,
derived for the range of radius covered by all sets of multiple images
shown in Figure~2, The profile is very flat within $\sim 200~kpc$, at a
level equal to the critical density for system~1 (black line). The
rise in the centre below $\sim 5"$ is due to the contribution of the
cD galaxy, under which the modelled smooth cluster mass component is
indicated by the black line. Three geometrically averaged Einstein
radii are indicated, corresponding to 16'' radius for system~1, and
27'' at the mean depth of the fainter systems ($z\simeq 2$) and 47''
is predicted for the Einstein radius at high redshift, $z\simeq8$.}
\end{inlinefigure}

\section{Brightest Cluster Galaxy}

The mass profiles of brightest cluster galaxies are not very well
constrained presently by lensing, but may help elucidate the origin of
these poorly understood class of objects (von der Linden et al., 2007,
and references therein). The best studied example with lensing is
MS2137 (Gavazzi et al., 2003, Donnarumma et al., 2009), where radial
arcs lie close to this object. For clusters like A1689 and Cl0024+17
there are too many massive galaxies in the central region to allow a
unique determination of the brightest galaxy. Here we are fortunate to
have such extended lensed images spread over a wide range of radius to
help constrain the central mass distribution. Furthermore, the
contrast of the cD mass above the flat cluster profile is clear,
allowing an accurate subtraction of the cluster contribution, as shown
in Figure 5. We model this galaxy analytically as a general power-law
mass profile with a projected core (Barkana 1998), resulting in a mass
of $\sim 1.0\pm 0.2 \times 10^{12} M_{\odot}$ interior to the limiting
radius of $\simeq 30~kpc$ traced by the low surface brightness wings
of the light profile (see Figure 1), where the uncertainly is
dominated by the gradient of the cD mass profile (Figure~5).

Integrating the light within this radius yields $2.2 \times 10^{11}
~L_{\odot}$, in the restframe B-band and hence a mean mass-to-light
ratio of $\sim M/L_B=4.5\pm 1~(M/L)_{\odot}$. This ratio can be fully accounted
for by the stars contained in this galaxy, for which we obtain
${M/L_B}\simeq5~(M/L)_{\odot}$, for a single burst stellar population formed
at $z=3$ and viewed at a redshift of $z=0.544$, equivalent to an age
of $\sim 6 ~Gyrs$, and with a mean metallicity of half solar. Other
lensing work supports relatively low mass-to-light for the cD galaxy
in MS2137 ($M/L <10~(M/L)_{\odot}$, Gavazzi et al., 2003), with less direct estimates relying on dynamical motions showing evidence for non-stellar DM at larger
radius for other cD galaxies (Dressler 1979, Gebhardt \& Thomas, 2009). This demonstrates
the need for more thorough studies of these enigmatic objects.

\section{Discussion}

The unusually large and undistorted lensed images of a spiral galaxy
uncovered here requires a nearly uniform distribution of matter within
the central $\sim 200~kpc$ region covered by these images. The
formation of multiple images requires the value of the central
surface density to be nearly equal to the critical surface density for lensing.
This lens corresponds to the case of maximum magnification, given by
the general relation $\mu=(((1-(\Sigma/\Sigma_{crit})^{2}-\gamma^{2}))^{-1}$, because the
lensing shear, $\gamma$, vanishes for a uniform density and therefore
the magnification diverges when the surface density is equal to the
critical density, $\Sigma=\Sigma_{crit}$. These conclusions follow
from the fundamentals of lensing and do not rely on any model.

In detail we show that the locations and parity of the lensed images
can be accurately reproduced with a simple model where indeed the
central mass distribution is nearly uniform and the magnification
derived is very large, $\times 200$, independent of the source
redshift. This model is used to identify 9 additional sets of fainter
multiply lensed images, which are incorporated to improve the fit, so
that the final model comprising 6 free parameters, is constrained by a
total of 33 lensed images. These fainter images are deflected by
larger angles than the bright lensed spiral galaxy placing their images
generally further from the cluster center.

From our model, the circular equivalent Einstein radius contained
within the critical area is $\simeq 27"$ for faint sources at $z\sim
2$, corresponding to the outer tangential critical curve (drawn in
Figure 2), which in physical units is $170\pm 20kpc$ at the distance
of the cluster. This very large radius adds to the already
uncomfortable discrepancy between the large Einstein radii observed
for massive clusters and the predictions based on the standard LCDM
cosmology (Broadhurst \& Barkana, 2008, Sadeh \& Rephaeli, 2008,
Puchwein \& Hilbert, 2009) for which such large Einstein radii can
only be contemplated with mass distributions which are prolate and
aligned along the line of sight (Corless \& King, 2007, Oguri \&
Blandford, 2009). Instead, here the cluster is evidently elongated
across the line of sight, traced by the bright galaxies, extended
X-ray emission (see Figure~1 in Ebeling et al., 2007) and the critical
curves (see Figure 2). The central mass distribution is evidently
unconcentrated, and presumably related to its currently unrelaxed
state.

The velocity field of the lensed spiral can be readily obtained
because of the large solid angle covered by each lensed image, and if
the inclination can be reliably derived from such data then an
independent estimation of the magnification can be derived via the
Tully-fisher relation. Spectroscopy of the internal velocity dispersion
of the starlight of the cD galaxy will also help in understanding the
cD mass profile, for which we have shown, rather surprisingly, seems to
be dominated by stars out to the observable limit of 30~kpc.

This cluster is unique in having near uniform density in projection, at
the critical level, thereby maximising gravitational lens
magnification. We calculate that the total area of sky exceeding a
magnification, $\mu>10$, is $\sim 2.8$ square arcminutes,
corresponding to the current high redshift limit of $z\sim 8$, which
is over twice the equivalent area calculated for other massive
clusters such as A1689 (Broadhurst et al., 2005) and Cl0024 (Zitrin et al., 2009a). This extreme magnification together with the lack
of image distortion makes MACS J1149.5+2223 the most powerful known
lens for accessing faint galaxies in the early universe.

\section*{acknowledgments}
We are grateful for the hospitality of ASIAA, where
part of this work was accomplished. We thank our anonymous referee for
useful comments, and acknowledge useful discussions with Craig
Sarazin, Holland Ford and Wei Zheng.  ACS was developed under NASA
contract NAS 5-32865. This research is based on observations provided
in the Hubble Legacy Archive which is a collaboration between the
Space Telescope Science Institute (STScI/NASA), the Space Telescope
European Coordinating Facility (ST-ECF/ESA) and the Canadian Astronomy
Data Centre (CADC/NRC/CSA).

\end{document}